\documentclass[prd,aps,showpacs,nofootinbib,preprint,eqsecnum]{revtex4}
\usepackage{graphicx,color,amsmath,amsxtra}
\usepackage{epsf}
\usepackage{amssymb}
\usepackage{enumerate}
\usepackage{hhline}
\usepackage{array}
\usepackage{tabularx}

\newcommand{\bear}{\begin{array}}  \newcommand{\eear}{\end{array}}
\newcommand{\bea}{\begin{eqnarray}}  \newcommand{\eea}{\end{eqnarray}}
\newcommand{\beq}{\begin{equation}}  \newcommand{\eeq}{\end{equation}}
\newcommand{\bef}{\begin{figure}}  \newcommand{\eef}{\end{figure}}
\newcommand{\bec}{\begin{center}}  \newcommand{\eec}{\end{center}}

\newcommand{\Eqn}[1]{&\hspace{-0.2em}#1\hspace{-0.2em}&}

\def\be{\begin{equation}}
\def\ee{\end{equation}}
\def\bea{\begin{eqnarray}}
\def\eea{\end{eqnarray}}
\def\beq{\begin{eqnarray}}
\def\eeq{\end{eqnarray}}

\def\be{\begin{equation}}
\def\ee{\end{equation}}
\def\bea{\begin{eqnarray}}
\def\eea{\end{eqnarray}}
\def\beq{\begin{eqnarray}}
\def\eeq{\end{eqnarray}}

\begin{document}

\title{Thermodynamics in $F(R)$ gravity with phantom crossing 
}

\author{Kazuharu Bamba\footnote{E-mail address: bamba``at"phys.nthu.edu.tw} 
and Chao-Qiang Geng\footnote{E-mail address: geng``at"phys.nthu.edu.tw}
}
\affiliation{
Department of Physics, National Tsing Hua University, Hsinchu, Taiwan 300
}


\begin{abstract}
We study thermodynamics of the apparent horizon in $F(R)$ gravity. 
In particular, we demonstrate that a $F(R)$ gravity model with realizing a 
crossing of the phantom divide can satisfy the second law of thermodynamics 
in the effective phantom phase as well as non-phantom one. 
\end{abstract}


\maketitle

\section{Introduction}

Recently, there have been more and more evidences to support 
that the current expansion of the universe 
is accelerating~\cite{WMAP1, SN1}. 
The scenarios to explain the current accelerated expansion of
the universe fall into two broad 
categories~\cite{Peebles:2002gy, Sahni:2005ct, 
Padmanabhan:2002ji, Copeland:2006wr, DM, Nojiri:2006ri, rv-2, 
Sotiriou:2008rp, Lobo:2008sg, Capozziello:2007ec}. 
One is to introduce ``dark energy'' in the framework of general relativity. 
The other is to study a modified gravitational theory, such as 
$F(R)$ gravity, in which the action is described by an arbitrary function 
$F(R)$ of the scalar curvature $R$ (for reviews, 
see~\cite{Nojiri:2006ri, rv-2, Sotiriou:2008rp, Lobo:2008sg, 
Capozziello:2007ec}). 

On the other hand, various observational data~\cite{observational status} 
imply that the ratio of the effective pressure to the effective energy of the 
universe, i.e., the effective equation of state (EoS) 
$w_\mathrm{eff}\equiv p_\mathrm{eff}/\rho_\mathrm{eff}$, 
may evolve from larger than $-1$ (non-phantom phase) to less 
than $-1$ (phantom phase~\cite{phantom}). Namely, it crosses $-1$ 
(the phantom divide) 
at the present time or in near future. 
Recently, an explicit model of $F(R)$ gravity with realizing 
a crossing of the phantom divide has been constructed in 
Ref.~\cite{Bamba:2008hq}. 
We note that the phantom crossing in the framework of general relativity has 
also been studied in the literature, e.g., ``quintom" model~\cite{quintom}. 

It is believed that a modified gravitational theory must pass cosmological 
bounds and solar 
system tests because it corresponds to an alternative theory of gravitation 
to general relativity. However, at the initial 
studies of $F(R)$ gravity, 
models proposed in Refs.~\cite{Capozziello, Carroll:2003wy, NO-i-s, 
Chiba:2003ir} with the powers of the scalar curvature are strongly 
constrained. In recent years, various investigations for viable models of 
$F(R)$ gravity~\cite{Hu:2007nk, Starobinsky:2007hu, Appleby:2007vb, 
Cognola:2007zu, Li:2007xn, Navarro:2006mw, NO-1, Amendola:2007nt, 
Tsujikawa:2007xu, Tsujikawa:2007tg, Amendola:2006we, Pogosian:2007sw, 
Faulkner:2006ub} have been executed. 
A curvature singularity problem in $F(R)$ gravity has also been discussed 
in Refs.~\cite{Frolov:2008uf, KM}.

As another touchstone of modified gravity, it is interesting 
to examine whether the second law of 
thermodynamics can be satisfied in the models of $F(R)$ gravity. 
The connection between gravitation and thermodynamics was 
examined by following black hole thermodynamics 
(black hole entropy~\cite{Bekenstein:1973ur} and 
temperature~\cite{Hawking:1974sw}) and its application to the cosmological 
event horizon of de Sitter space~\cite{Gibbons:1977mu}. 
It was shown that Einstein equation is derived from the proportionality of 
the entropy to the horizon area together with the fundamental thermodynamic 
relation, such as the Clausius relation~\cite{Jacobson:1995ab}. 
This idea has been employed in a cosmological 
context~\cite{Frolov:2002va, Danielsson:2004xw, Bousso:2004tv, Cai:2005ra}. 
It was demonstrated that if the entropy of the apparent horizon in the 
Friedmann-Robertson-Walker (FRW) spacetime is proportional to the apparent 
horizon area, Friedmann equations follow from the first law of 
thermodynamics~\cite{Cai:2005ra}. 
The equivalent considerations for the FRW universe with the viscous fluid have 
also been studied~\cite{Akbar:2008vc}. 

In addition, it was proposed~\cite{Eling:2006aw} that in $F(R)$ gravity, 
a non-equilibrium thermodynamic treatment should be required 
in order to derive the corresponding gravitational field equation by using the 
procedure in Ref.~\cite{Jacobson:1995ab}. 
It was reconfirmed in Ref.~\cite{Akbar:2006mq} in $F(R)$ gravity as well as 
Ref.~\cite{Cai:2006rs} in scalar-tensor theories. 
The first~\cite{Wu:2007se} and second~\cite{Wu:2008ir} laws of 
thermodynamics on the apparent horizon in generalized theories of gravitation 
have recently been analyzed by taking into account the non-equilibrium 
thermodynamic treatment. 
Reinterpretations of the non-equilibrium correction~\cite{Eling:2006aw} 
through the introduction of a mass-like function~\cite{Gong:2007md} and other 
approaches~\cite{Eling:2008af,Elizalde:2008pv} 
have also been explored. 
Incidentally, the horizon entropy in four-dimensional 
modified gravity~\cite{Wang:2005bi} and a quantum logarithmic correction to 
the expression of the horizon entropy in a cosmological 
context~\cite{Cai:2008ys, Lidsey:2008zq, Zhu:2008cg} have been examined. 
Moreover, the first law of the ordinary equilibrium thermodynamics in $F(R)$ 
gravity, scalar-tensor theories, the 
Gauss-Bonnet gravity and more general Lovelock gravity have been discussed in
Refs.~\cite{Akbar:2006er,Akbar:2006kj,Cai:2008mh, Paranjape:2006ca}, while 
the corresponding studies on the second law in the accelerating universe, 
$F(R)$ gravity, the Gauss-Bonnet gravity 
and the Lovelock gravity have been done in 
Refs.~\cite{Zhou:2007pz,MohseniSadjadi:2007zq,Akbar:2008vz}, respectively. 
Studies of thermodynamics in braneworld scenario~\cite{BW-1, BW-2, Zhu:2008di} 
as well as its properties of dark energy~\cite{Saridakis:2009uu} have also 
been performed. 

It was pointed out in Ref.~\cite{Brevik:2004sd} that thermodynamics in the 
phantom phase usually leads to a negative entropy. Moreover, it was 
noted~\cite{Buchmuller:2006em} that in the framework of general relativity the 
horizon entropy decreases in phantom models. However, the conditions that 
the black hole entropy can be positive in the $F(R)$ gravity 
models~\cite{Hu:2007nk, Starobinsky:2007hu, Appleby:2007vb} with the 
solar-system tests have been analyzed in Ref.~\cite{Briscese:2007cd}. 
These recent studies have motivated us to explore whether in the framework of 
$F(R)$ gravity the second law of thermodynamics can be satisfied in the 
phantom phase. To illustrate the point, in the present paper we consider a 
$F(R)$ gravity model with a crossing of the phantom divide~\cite{Bamba:2008hq} 
as it contains an effective phantom phase. 
We use units of $k_\mathrm{B} = c = \hbar = 1$ and denote the
gravitational constant $8 \pi G$ by 
${\kappa}^2 \equiv 8\pi/{M_{\mathrm{Pl}}}^2$ 
with the Planck mass of $M_{\mathrm{Pl}} = G^{-1/2} = 1.2 \times 10^{19}$GeV.

The paper is organized as follows. 
In Sec.\ II, we explain the first and second laws of thermodynamics 
in $F(R)$ gravity. 
In Sec.\ III, we demonstrate that the model of 
$F(R)$ gravity with the phantom crossing~\cite{Bamba:2008hq} 
can satisfy the generalized second law of the thermodynamics. 
Finally, conclusions are given in Sec.\ IV.

\section{Thermodynamics in $F(R)$ gravity}

In this section, we study the first and second laws of thermodynamics of the 
apparent horizon in $F(R)$ gravity. 
We consider the four-dimensional flat spacetime.

\subsection{$F(R)$ gravity}

The action of $F(R)$ gravity with matter is written as
\begin{eqnarray}
I = \int d^4 x \sqrt{-g} \left[ \frac{F(R)}{2\kappa^2} +
{\mathcal{L}}_{\mathrm{matter}} \right]\,,
\label{eq:2.1}
\end{eqnarray}
where $g$ is the determinant of the metric tensor $g_{\mu\nu}$ and
${\mathcal{L}}_{\mathrm{matter}}$ is the matter Lagrangian. 
  From the action in Eq.~(\ref{eq:2.1}), the field equation of modified 
gravity is given by 
\begin{eqnarray}
F^{\prime}(R) R_{\mu \nu}
- \frac{1}{2}g_{\mu \nu} F(R) + g_{\mu \nu}
\Box F^{\prime}(R) - {\nabla}_{\mu} {\nabla}_{\nu} F^{\prime}(R)
= \kappa^2 T^{(\mathrm{matter})}_{\mu \nu}\,, 
\label{eq:2.2}
\end{eqnarray}
where the prime denotes differentiation with respect to $R$, 
${\nabla}_{\mu}$ is the covariant derivative operator associated with 
$g_{\mu \nu}$, $\Box \equiv g^{\mu \nu} {\nabla}_{\mu} {\nabla}_{\nu}$
is the covariant d'Alembertian for a scalar field, 
$R_{\mu \nu}$ is the Ricci curvature tensor, 
and 
$T^{(\mathrm{matter})}_{\mu \nu} = \mathrm{diag} \left(\rho, p, p, p \right)$ 
is the contribution to the energy-momentum tensor from 
all ordinary matters
with $\rho$ and $p$ being the energy density and pressure of all ordinary 
matters, respectively. 

We assume the flat 
Friedmann-Robertson-Walker (FRW) space-time with the metric,
\begin{eqnarray}
{ds}^2 \Eqn{=} -{dt}^2 + a^2(t)\gamma_{ij}dx^i dx^j,  
\label{eq:2.3} \\ 
\gamma_{ij}dx^i dx^j \Eqn{=} dr^2 + r^2 d\Omega^2\,,
\label{eq:2.4}
\end{eqnarray}
where $a(t)$ is the scale factor and $d\Omega^2$ is the metric of 
two-dimensional sphere with unit radius. 
In the FRW background, from $(\mu,\nu)=(0,0)$ 
and the trace part of $(\mu,\nu)=(i,j)$ $(i,j=1,\cdots,3)$ components in 
Eq.~(\ref{eq:2.2}), we obtain the gravitational field equations: 
\begin{eqnarray}
H^2 \Eqn{=} \frac{\kappa^2}{3F^{\prime}(R)} \left( \rho + \rho_\mathrm{c} 
\right)\,,
\label{eq:2.5} \\ 
\dot{H} \Eqn{=} -\frac{\kappa^2}{2F^{\prime}(R)} \left( \rho + p + 
\rho_\mathrm{c} + p_\mathrm{c}
\right)
\,,
\label{eq:2.6}
\end{eqnarray} 
where $\rho_\mathrm{c}$ and $p_\mathrm{c}$ can be 
regarded as the energy density and pressure generated due to the difference of 
$F(R)$ gravity from general relativity, 
given by 
\begin{eqnarray} 
\rho_\mathrm{c} \Eqn{=} \frac{1}{\kappa^2} \left[ \frac{1}{2} \left( 
-F(R) +RF^{\prime}(R) \right) -3H\dot{R}F^{\prime\prime}(R) \right]\,, 
\label{eq:2.7} \\
p_\mathrm{c} \Eqn{=} \frac{1}{\kappa^2} \left[ \frac{1}{2} \left( 
F(R) - RF^{\prime}(R) \right) + \left(2H\dot{R} + \ddot{R} \right) 
F^{\prime\prime}(R) + \dot{R}^2 F^{\prime\prime\prime}(R) \right]\,,
\label{eq:2.8}
\end{eqnarray} 
respectively,
with the scalar curvature of $R=6\left(\dot{H} + 2H^2 \right)$. 
Here, $H=\dot{a}/a$ is the Hubble parameter and 
the dot denotes the time derivative of $\partial/\partial t$. 
We define the effective energy density and pressure of the universe as 
$\rho_\mathrm{eff} \equiv \rho_\mathrm{t}/F^{\prime}(R)$ and 
$p_\mathrm{eff} \equiv p_\mathrm{t}/F^{\prime}(R)$
with $\rho_\mathrm{t} = \rho+\rho_\mathrm{c}$ and $p_\mathrm{t} = 
p+p_\mathrm{c}$, respectively. 
Hence, from Eqs.~(\ref{eq:2.5}) and (\ref{eq:2.6}) we see that 
even in $F(R)$ gravity, the gravitational field equations are expressed as 
$H^2 = \kappa^2 \rho_\mathrm{eff} /3$ and 
$\dot{H} =  -\kappa^2 \left( \rho_\mathrm{eff} + p_\mathrm{eff} \right)/2$, 
which are the same  as those in general relativity. 

The continuity equation in terms of the effective energy density and pressure 
of the universe is given by 
\begin{eqnarray} 
\dot{\rho}_\mathrm{eff} + 3H\left( \rho_\mathrm{eff} + p_\mathrm{eff} \right) 
= 0\,.
\label{eq:2.9}
\end{eqnarray} 
Similarly, the (semi-)continuity equation of ordinary matters has the form 
\begin{eqnarray} 
\dot{\rho} + 3H\left( \rho + p \right) = q\,.
\label{eq:2.10}
\end{eqnarray}
One can take $q=0$ because the gravity is determined only by ordinary matters. 
Assuming that the energy fluid, generated from the modification of gravity, 
behaves as a perfect fluid, we have similar semi-continuity equations as 
\begin{eqnarray} 
\dot{\rho}_\mathrm{c} + 3H\left( \rho_\mathrm{c} + p_\mathrm{c} \right) 
\Eqn{=} q_\mathrm{c}\,,
\label{eq:2.11} \\
\dot{\rho}_\mathrm{t} + 3H\left( \rho_\mathrm{t} + p_\mathrm{t} \right) 
\Eqn{=} q_\mathrm{t}\,,
\label{eq:2.12}
\end{eqnarray} 
where $q_\mathrm{c}$ and $q_\mathrm{t} (=q+q_\mathrm{c})$ are quantities 
of expressing energy exchange. 
Using Eqs.~(\ref{eq:2.5}), (\ref{eq:2.6}) and (\ref{eq:2.12}), we obtain 
\begin{eqnarray} 
q_\mathrm{t} = \frac{3}{\kappa^2} H^2 
\frac{\partial F^{\prime}(R)}{\partial t}\,.
\label{eq:2.13}
\end{eqnarray} 
Clearly, from Eq.~(\ref{eq:2.13}), we find that $q_\mathrm{t} = 0$ 
for general relativity with $F(R) = R$, whereas 
$q_\mathrm{t}$ does not generally vanish in $F(R)$ gravity 
since there could exist some energy exchange with the horizon.

\subsection{First law of thermodynamics}

We now illustrate the first law of thermodynamics in $F(R)$ gravity. 
By using the spherical symmetry, the metric~(\ref{eq:2.3}) can be written as 
\begin{eqnarray}
{ds}^2 = h_{\alpha \beta}dx^{\alpha} dx^{\beta} + \tilde{r}^2d\Omega^2\,, 
\label{eq:2.14}
\end{eqnarray}
where $\tilde{r}=a(t)r$, $x^0 =t$ and $x^1=r$, and $h_{\alpha \beta}$ is 
the two-dimensional metric $h_{\alpha \beta} = \mathrm{diag} 
\left(-1,a^2\right)$. The dynamical apparent horizon is determined by the 
relation $h^{\alpha\beta} \partial_\alpha \tilde{r} \partial_\beta \tilde{r} 
= 0$. The radius of the apparent horizon for the FRW spacetime is given 
by~\cite{Cai:2005ra, Wu:2008ir} 
\begin{eqnarray} 
\tilde{r}_A =\frac{1}{H}\,. 
\label{eq:2.15}
\end{eqnarray}
The associated temperature $T$ of the apparent horizon is determined through 
the surface gravity of 
\begin{eqnarray} 
\kappa_{\mathrm{sg}} = \frac{1}{2\sqrt{-h}} \partial_\alpha 
\left( \sqrt{-h}h^{\alpha\beta} \partial_\beta \tilde{r} \right)\,, 
\label{eq:2.16}
\end{eqnarray}
where $h$ is the determinant of the metric $h_{\alpha\beta}$. 
We note that the recent type Ia Supernovae data suggests 
that in the accelerating universe the enveloping surface should 
be the apparent horizon rather than the event one from the thermodynamic point 
of view~\cite{Zhou:2007pz}. 

In the FRW spacetime, one has~\cite{Cai:2005ra} 
\begin{eqnarray} 
T=\frac{|\kappa_{\mathrm{sg}}|}{2\pi} = 
\frac{1}{2\pi \tilde{r}_A} \left(1-\frac{\dot{\tilde{r}}_A}{2H\tilde{r}_A} 
\right)\,.
\label{eq:2.18}
\end{eqnarray} 
In general relativity, the entropy is expressed as 
\begin{eqnarray}  
S_\mathrm{GR} = \frac{A}{4G}\,, 
\label{eq:2.19}
\end{eqnarray}
where $A=4\pi \tilde{r}_A^2$ is the horizon area. 
It follows from Eqs.~(\ref{eq:2.18}) and (\ref{eq:2.19}) that 
\begin{eqnarray}  
TdS_\mathrm{GR} 
\Eqn{=} 
\frac{1}{G} \left( 1-\frac{\dot{\tilde{r}}_A}{2H\tilde{r}_A} \right) 
d\tilde{r}_A \nonumber \\
\Eqn{=} -\frac{3V}{\kappa^2} \frac{dH^2}{dt}dt -\frac{3V}{\kappa^2} 
\frac{\dot{H}^2}{H}dt\,,
\label{eq:2.20}
\end{eqnarray}
where $V = 4\pi\tilde{r}_A^3/3$ is the volume of the apparent horizon. 
 From Eq.~(\ref{eq:2.5}), we see that 
$H^2 = H^2\left( \rho, \rho_\mathrm{c}, F^{\prime}(R) \right)$ and hence 
$
dH^2/dt = 
\left( \partial H^2/\partial \rho \right) \dot{\rho} + 
\left( \partial H^2/\partial \rho_\mathrm{c} \right) 
\dot{\rho}_\mathrm{c} + 
\left( \partial H^2/\partial F^{\prime}(R) \right) 
\dot{F}^{\prime}(R)
$. 
Substituting this equation into Eq.~(\ref{eq:2.20}) and multiplying 
the resultant equation by the factor $\left(\kappa^2/3\right) 
\left( \partial H^2/\partial \rho \right)^{-1}$, we get
\begin{eqnarray}
\hspace{-8mm}
\frac{\kappa^2}{3} \left(\frac{\partial H^2}{\partial \rho}\right)^{-1} T 
dS_\mathrm{GR} \Eqn{=} -V \dot{\rho}dt -V 
\left(\frac{\partial H^2}{\partial \rho}\right)^{-1} \frac{\dot{H}^2}{H} dt 
- V \left(\frac{\partial H^2}{\partial \rho}\right)^{-1} 
\left(\frac{\partial H^2}{\partial \rho_\mathrm{c}}\right) 
\dot{\rho}_\mathrm{c} dt 
\nonumber \\
&& 
- V \left(\frac{\partial H^2}{\partial \rho}\right)^{-1} 
\left(\frac{\partial H^2}{\partial F^{\prime}(R)} \right) 
\dot{F}^{\prime}(R) dt\,. 
\label{eq:2.21}
\end{eqnarray}
The left-hand side of Eq.~(\ref{eq:2.21}) can be rewritten as
\begin{eqnarray}
\frac{\kappa^2}{3} \left(\frac{\partial H^2}{\partial \rho}\right)^{-1} T 
dS_\mathrm{GR} = 
T d\left[ \frac{\kappa^2}{3} \left(\frac{\partial H^2}{\partial \rho} 
\right)^{-1} S_\mathrm{GR} \right] - T \frac{\kappa^2}{3} S_\mathrm{GR} 
d\left[ \left(\frac{\partial H^2}{\partial \rho} 
\right)^{-1} \right]\,,
\label{eq:2.22}
\end{eqnarray} 
which leads to the Clausius relation
\begin{eqnarray}
TdS=\delta Q\,, 
\label{eq:2.23}
\end{eqnarray}
where the entropy $S$ and the energy flux $\delta Q$ are defined by
\begin{eqnarray}
S \Eqn{\equiv} \frac{\kappa^2}{3} \left(\frac{\partial H^2}{\partial \rho} 
\right)^{-1} S_\mathrm{GR}\,, 
\label{eq:2.24} \\
\delta Q \Eqn{\equiv} 
-V \dot{\rho}dt -V 
\left(\frac{\partial H^2}{\partial \rho}\right)^{-1} \frac{\dot{H}^2}{H} dt 
- V \left(\frac{\partial H^2}{\partial \rho}\right)^{-1} 
\left(\frac{\partial H^2}{\partial \rho_\mathrm{c}}\right) 
\dot{\rho}_\mathrm{c} dt 
\nonumber \\
&&
{}+T \frac{\kappa^2}{3} S_\mathrm{GR} 
d\left[ \left(\frac{\partial H^2}{\partial \rho} \right)^{-1} \right]
- V \left(\frac{\partial H^2}{\partial \rho}\right)^{-1} 
\left(\frac{\partial H^2}{\partial F^{\prime}(R)} \right) 
\dot{F}^{\prime}(R) dt\,.
\label{eq:2.25}
\end{eqnarray} 
 From Eqs.~(\ref{eq:2.5}), (\ref{eq:2.19}) and (\ref{eq:2.24}), 
the entropy in $F(R)$ gravity is expressed 
as~\cite{entropy-1, Brevik:2004sd, Cognola:2005de}:
\begin{eqnarray}
S=\frac{AF^{\prime}(R)}{4G}\,.
\label{eq:Entropy}
\end{eqnarray} 
We note that there is arbitrariness in the definitions of 
$S$ and $\delta Q$. The reason for using those in 
Eqs.~(\ref{eq:2.24}) and (\ref{eq:2.25}) is that 
we can obtain the simple form of Eq.~(\ref{eq:Entropy}) as 
the expression of the entropy in $F(R)$ gravity.
Using $\rho_\mathrm{eff} = 3H^2/\kappa^2$ and Eq.~(\ref{eq:2.5}), we find
\begin{eqnarray}
\dot{\rho}_\mathrm{eff} \Eqn{=} \frac{3}{\kappa^2} 
\left[ \left(\frac{\partial H^2}{\partial \rho}\right) \dot{\rho} 
+\left(\frac{\partial H^2}{\partial \rho_\mathrm{c}}\right) 
\dot{\rho}_\mathrm{c} 
+\left(\frac{\partial H^2}{\partial F^{\prime}(R)} \right) 
\dot{F}^{\prime}(R) \right] 
\nonumber \\
\Eqn{=} \frac{\dot{\rho}}{F^{\prime}(R)} 
+\frac{3}{\kappa^2} 
\left[ 
\left(\frac{\partial H^2}{\partial \rho_\mathrm{c}}\right) 
\dot{\rho}_\mathrm{c}
+\left(\frac{\partial H^2}{\partial F^{\prime}(R)} \right) 
\dot{F}^{\prime}(R) \right]\,, 
\label{eq:2.26} \\
\rho_\mathrm{eff} + p_\mathrm{eff} \Eqn{=} 
\frac{\rho_\mathrm{t} + p_\mathrm{t}}{ F^{\prime}(R) }\,, 
\label{eq:2.I-1}
\end{eqnarray}
where we have used 
$\left( \partial H^2/\partial \rho \right) = \kappa^2/\left(3F^{\prime}(R)
\right)$. 
By combining Eqs.~(\ref{eq:2.9}), (\ref{eq:2.10}) and (\ref{eq:2.26}), 
we obtain 
\begin{eqnarray}
\left(\frac{\partial H^2}{\partial \rho_\mathrm{c}}\right) 
\dot{\rho}_\mathrm{c} 
+\left(\frac{\partial H^2}{\partial F^{\prime}(R)} \right) 
\dot{F}^{\prime}(R)
= -\frac{\kappa^2}{3F^{\prime}(R)}\left[ 
3H\left( \rho_\mathrm{t}  + p_\mathrm{t} - \rho -p \right) + q
\right]\,.
\label{eq:2.27}
\end{eqnarray}
Consequently, from Eqs.~(\ref{eq:2.6}), (\ref{eq:2.10}), (\ref{eq:2.25}) and 
(\ref{eq:2.27}), we get
\begin{eqnarray}
\delta Q \Eqn{=} 3H\left( \rho_\mathrm{t} + p_\mathrm{t} \right)Vdt 
-\frac{1}{2}\left( \rho_\mathrm{t} + p_\mathrm{t} \right)\dot{V}dt 
+TS_\mathrm{GR}d\left( F^{\prime}(R) \right)
\label{eq:2.28} \\
\Eqn{=} -dE_\mathrm{t} +W_\mathrm{t}dV +q_\mathrm{t}Vdt +
TS_\mathrm{GR}d\left( F^{\prime}(R) \right)\,,
\label{eq:2.29}
\end{eqnarray}
where $E_\mathrm{t} = \rho_\mathrm{t} V$ is the total intrinsic energy and 
$W_\mathrm{t} \equiv -\left(1/2\right) T^{(\mathrm{t})\alpha\beta}
h_{\alpha\beta} 
= \left( \rho_\mathrm{t} - p_\mathrm{t} \right)/2$ is the total 
work density~\cite{work density}. 
Here, $T^{(\mathrm{t})}_{\mu\nu} = \mathrm{diag} \left(\rho_\mathrm{t}, 
p_\mathrm{t}, p_\mathrm{t}, p_\mathrm{t} \right)$ is 
the contribution to the energy-momentum tensor from 
all ordinary matters and energy fluid. 
This may be regarded as the work generated through the evolution of 
the apparent horizon~\cite{work density}. 
It follows from Eqs.~(\ref{eq:2.18}), (\ref{eq:2.19}) and (\ref{eq:2.29}) that the first law of thermodynamics 
in modified gravity can be constructed as 
\begin{eqnarray}
\delta Q = -dE_\mathrm{t} +W_\mathrm{t}dV -Td_\mathrm{c}S\,, 
\label{eq:2.30}
\end{eqnarray}
where
\begin{eqnarray}
d_\mathrm{c}S \Eqn{=} -\frac{1}{T}q_\mathrm{t}Vdt -S_\mathrm{GR}
d\left( F^{\prime}(R) \right) 
\label{eq:2.31} \\
\Eqn{=} -\frac{8\pi^2}{\kappa^2} \frac{4H^2+\dot{H}}{\left(2H^2
+\dot{H}\right)H^2} d\left( F^{\prime}(R) \right)\,.
\label{eq:2.32}
\end{eqnarray} 
 From Eq.~(\ref{eq:2.32}), it is clear that $d_\mathrm{c}S$ 
does not vanish if $F(R)$ is not equal to $R$. 
Hence, the emergence of the entropy production term is a 
special feature of thermodynamics in $F(R)$ gravity. We note that such 
feature also appears in scalar-tensor theories because 
in those theories $F^{\prime}(R)$ is not constant but some dynamical 
quantities in terms of scalar fields. 
We remark that there exists a possible singularity in Eq.~(\ref{eq:2.32})
if $2H^2+\dot{H} = 0$, which corresponds to $T=0$ since 
$T=\left( 2H^2+\dot{H} \right)/\left( 4\pi H \right)$ due to 
Eqs.~(\ref{eq:2.15}) and (\ref{eq:2.18}). 
It is clear that the necessary condition for a positive temperature is 
$2H^2+\dot{H} >0$. In this case, no singularity appears in 
Eq.~(\ref{eq:2.32}). 

Using the Clausius relation in Eq.~(\ref{eq:2.23}), 
the first law of thermodynamics in $F(R)$ gravity in Eq. (\ref{eq:2.30}) can 
be rewritten as~\cite{Wu:2008ir}
\begin{eqnarray}
TdS + Td_\mathrm{c}S = -dE_\mathrm{t} +W_\mathrm{t}dV\,,
\label{eq:2.33}
\end{eqnarray} 
which characterizes the non-equilibrium 
thermodynamics of the apparent horizon in $F(R)$ gravity.

\subsection{Second law of thermodynamics}

Next, we investigate the second law of thermodynamics in $F(R)$ gravity. From 
Eq.~(\ref{eq:2.33}), the first law of thermodynamics in 
terms of the horizon entropy $S_\mathrm{h}$ is expressed as 
\begin{eqnarray}
TdS_\mathrm{h} = -dE_\mathrm{t} +W_\mathrm{t}dV -Td_\mathrm{c}S\,.
\label{eq:2.34}
\end{eqnarray}
The Gibbs equation in terms of all matter and energy fluid is given by 
\begin{eqnarray}
T_\mathrm{t} dS_\mathrm{t} = d\left(\rho_\mathrm{t}V\right) + p_\mathrm{t}dV 
=Vd\rho_\mathrm{t} + \left( \rho_\mathrm{t} + p_\mathrm{t} \right)dV\,, 
\label{eq:2.35}
\end{eqnarray}
where $T_\mathrm{t}$ and $S_\mathrm{t}$ denote the temperature and entropy of 
total energy inside the horizon, respectively. We assume that 
\begin{eqnarray}
T_\mathrm{t} = b T\,,
\label{eq:2.36}
\end{eqnarray}
where $b$ is a constant with $0<b<1$. If there is no energy 
exchange between the outside and inside of the apparent horizon, 
i.e., $q_\mathrm{t} = 0$, thermal equilibrium realizes and 
therefore $b=1$. 

The second law of thermodynamics in $F(R)$ gravity can be described 
by~\cite{Wu:2008ir}
\begin{eqnarray} 
\dot{S}_\mathrm{h} + \frac{\partial \left( d_\mathrm{c}S \right)}{\partial t} 
+ \dot{S}_\mathrm{t} \geq 0\,
\label{eq:2.37}
\end{eqnarray}
or 
\begin{eqnarray} 
\left(1-b\right) \dot{\rho}_\mathrm{t} V + \left(1-\frac{b}{2}\right) 
\left( \rho_\mathrm{t} + p_\mathrm{t} \right) \dot{V} \geq 0\,
\label{eq:2.38}
\end{eqnarray}
by using Eqs.~(\ref{eq:2.34})--(\ref{eq:2.36}). 
With Eqs.~(\ref{eq:2.5}) and (\ref{eq:2.6}), 
the relation in Eq. (\ref{eq:2.38}) is reduced to 
\begin{eqnarray}
\frac{4\pi}{\kappa^2} \frac{1}{H^4} J \geq 0\,, 
\label{eq:2.39} 
\end{eqnarray}
where 
\begin{eqnarray}
J = \left(1-b\right) H^3 \dot{R}F^{\prime\prime}(R) + 2 \left(1-b\right) 
H^2\dot{H}F^{\prime}(R) + \left(2-b\right)\dot{H}^2F^{\prime}(R)\,. 
\label{eq:2.40} 
\end{eqnarray}
Thus, the condition to satisfy the second law of thermodynamics in $F(R)$ 
gravity is equivalent to $J \geq 0$. 

We note that for $F(R) = R$, in which 
$q_\mathrm{t} = 0$, $b=1$ and $d_\mathrm{c}S =0$ with the realization of 
thermal equilibrium, we find $J=\dot{H}^2$ from Eq.~(\ref{eq:2.40}). 
In general relativity with a pure de Sitter expansion ($\dot{H}=0$), 
in which $F(R)$ is given by 
$F(R) = R-2\Lambda$ with $\Lambda$ being the cosmological constant, 
$J=0$.
It is clear that $J$ does not vanish for $b=1$ if $\dot{H}\neq 0$. 

\section{Second law of thermodynamics in a $F(R)$ gravity model realizing a 
crossing of the phantom divide}

In this section, we examine whether a $F(R)$ gravity model~\cite{Bamba:2008hq} 
with the phantom crossing can satisfy the second law of thermodynamics 
discussed in Sec.~II. In the model~\cite{Bamba:2008hq}, the Hubble rate $H(t)$ 
is given by 
\begin{eqnarray}
H(t) = \left(\frac{10}{t}\right) \left[\frac{\gamma + (\gamma+1) 
\left(\frac{t}{t_s}\right)^{2\gamma+1}}{1 - 
\left(\frac{t}{t_s}\right)^{2\gamma+1}}\right]\,,
\label{eq:3.1} 
\end{eqnarray}
where $\gamma$ is a positive constant and $t_s$ is the time when the Big Rip 
singularity appears.
Here, we only consider the period $0<t<t_s$. 
When $t\to 0$, i.e., $t \ll t_s$, $H(t)$ behaves as 
\begin{eqnarray}
H(t) \sim \frac{10\gamma}{t}\,.
\label{eq:3.2} 
\end{eqnarray} 
In the FRW background~(\ref{eq:2.3}), the effective EoS 
$w_\mathrm{eff}$ is given by~\cite{Nojiri:2006ri} 
$w_\mathrm{eff} 
= -1 -2\dot{H}/\left(3H^2\right)$. 
In the limit of $t\to 0$, 
$w_\mathrm{eff} = -1 + 1/\left(15\gamma\right) > -1$, 
corresponding to the non-phantom phase. 
The form of $F(R)$ is given by 
\begin{eqnarray}
F(R) \Eqn{\sim}
\left\{
\frac{\left[\frac{1}{t_0} \sqrt{60\gamma \left( 20\gamma -1 \right)} R^{-1/2}
\right]^\gamma}{1 -
\left[\frac{1}{t_s} \sqrt{60\gamma \left( 20\gamma -1 \right)} 
R^{-1/2}
\right]^{2\gamma+1}} \right\}^5 R \nonumber \\
&& \hspace{10mm}
{}\times
\sum_{j=\pm}
\biggl\{ \left( \frac{5\gamma -\beta_j -1}{20\gamma -1} \right) \tilde{p}_j
\left[60\gamma \left( 20\gamma -1 \right) \right]^{\beta_j /2}
R^{-\beta_j /2}
\biggr\}\,,
\label{eq:3.3}
\end{eqnarray}
where 
\begin{eqnarray}
\beta_\pm = \frac{1 \pm \sqrt{1+100\gamma \left(\gamma+1\right)}}{2}\,. 
\label{eq:3.4}
\end{eqnarray} 
Here, $t_0$ is the present time and $\tilde{p}_\pm$ are arbitrary constants. 
We remark that the stability for the obtained solutions in Eq.~(\ref{eq:3.3}) 
under a quantum correction coming from conformal anomaly has been examined 
in Ref.~\cite{Bamba:2008hq}. 
It has been shown that the quantum correction could 
be small when the phantom crossing occurs, although it becomes important near 
the Big Rip singularity. 

Using Eq.~(\ref{eq:3.2}), we find 
$R \sim 60\gamma \left( 20\gamma -1 \right)/t^2$ 
and $\sqrt{60\gamma \left( 20\gamma -1 \right)} 
R^{-1/2}/t_s \sim t/t_s \ll 1$. 
As the denominator inside 
the first large braces $\{\,\}$ on the right-hand side of Eq.~(\ref{eq:3.3}) 
is approximately unity, 
Eq.~(\ref{eq:3.3}) can be simplified to 
\begin{eqnarray}
F(R) \Eqn{\approx} 
\left[ \frac{1}{t_0} \sqrt{60\gamma \left( 20\gamma -1 \right)} 
\right]^{5\gamma} R^{-5\gamma/2 +1} 
\nonumber \\
&& \hspace{10mm}
{}\times
\sum_{j=\pm}
\left\{ \left( \frac{5\gamma -\beta_j -1}{20\gamma -1} \right) \tilde{p}_j
\left[60\gamma \left( 20\gamma -1 \right) \right]^{\beta_j /2}
R^{-\beta_j /2}
\right\}\,.
\label{eq:3.5}
\end{eqnarray} 
In this case, we obtain
\begin{eqnarray}
J \Eqn{=} 
50\gamma^2 \left[ \frac{1}{t_0} \sqrt{60\gamma \left( 20\gamma -1 \right)} 
\right]^{5\gamma} \frac{1}{t^4}  R^{-5\gamma/2} 
\sum_{j=\pm} 
\biggl\{ \left[ 10\gamma\left(1-b\right) \left(-5\gamma-\beta_j+2\right) - 
\left(2-b\right) \right] 
\nonumber \\
&& \hspace{20mm}
{}\times
\left( \frac{5\gamma -\beta_j -1}{20\gamma -1} \right) \tilde{p}_j 
\left[60\gamma \left( 20\gamma -1 \right) \right]^{\beta_j /2} 
\left(5\gamma+\beta_j-2\right) R^{-\beta_j /2}
\biggr\}\,.
\label{eq:3.6}
\end{eqnarray}
 Note that $R>0$ for $R \sim 60\gamma \left( 20\gamma -1 \right)/t^2$ and 
$\gamma > 1/20$. 
Using Eqs.~(\ref{eq:2.40}), (\ref{eq:3.4}) and (\ref{eq:3.6}) and taking into 
account the fact that $\tilde{p}_\pm$ are arbitrary constants, 
the necessary condition to have $J \geq 0$ is 
\begin{eqnarray}
\tilde{p}_\pm \left\{ 5\gamma\left(1-b\right) \left[ \left(10\gamma-3\right) 
\pm \sqrt{1+100\gamma \left(\gamma+1\right)} \right] +2-b \right\}
\geq 0\,,
\label{eq:3.8}
\end{eqnarray}
where we have assumed $\gamma > 1/20$. 
We remark that for simplicity, we have chosen positive 
coefficients of $R^{-\beta_+ /2}$ and $R^{-\beta_- /2}$ in 
the large braces $\{\,\}$ of Eq.~(\ref{eq:3.6}). 
By taking the values of $\tilde{p}_\pm$ so that the relation 
(\ref{eq:3.8}) can be met, the second law of thermodynamics, 
i.e., $J \geq 0$, can be satisfied. 

It follows from $w_\mathrm{eff} = -1 -2\dot{H}/\left(3H^2\right)$ that 
$w_\mathrm{eff}=-1$ when $\dot{H}=0$.
Solving $w_\mathrm{eff} = -1$ with respect to $t$ by using 
Eq.~(\ref{eq:3.1}), we find that 
$w_\mathrm{eff}$ crosses the 
phantom divide at the time $t = t_\mathrm{c}$, given by
\begin{eqnarray}
t_\mathrm{c} = t_s \left( -2\gamma +
\sqrt{4\gamma^2 + \frac{\gamma}{\gamma+1}}
\right)^{1/\left( 2\gamma + 1 \right)}\,.
\label{eq:3.9}
\end{eqnarray}
On the other hand, when $t\to t_s$, we have 
\begin{eqnarray}
H(t) \sim \frac{10}{t_s - t}\,.
\label{eq:3.10}
\end{eqnarray} 
In this case, the scale factor is given by 
$a(t) \sim \bar{a} \left( t_s - t \right)^{-10}$ 
with a constant of $\bar{a}$. When $t\to t_s$, $a \to \infty$ and therefore the Big Rip singularity appears. 
In this limit, $w_\mathrm{eff} = -16/15 < -1$, 
corresponding to the phantom phase. The form of $F(R)$ is given by 
\begin{eqnarray}
F(R) \Eqn{\sim}
\left(
\frac{
\left\{ \frac{1}{t_0}
\left[ t_s - 3\sqrt{140} R^{-1/2} \right]
\right\}^\gamma}
{1 -
\left[ 1 - \frac{3\sqrt{140}}{t_s} R^{-1/2}
\right]^{2\gamma+1}
} \right)^5 R
\sum_{j=\pm}
\tilde{p}_j
\left[ t_s - 3\sqrt{140} R^{-1/2}
\right]^{\beta_j}
\nonumber \\
&& \hspace{0mm}
{}\times
\Biggl\{
1- \sqrt{\frac{20}{7}}
\left[
\sqrt{\frac{15}{84}} t_s
+ \left( \beta_j - 15 \right) R^{-1/2}
\right]
\frac{1}{t_s - 3\sqrt{140} R^{-1/2}}
\Biggr\}\,,
\label{eq:3.11}
\end{eqnarray} 
which is reduced to 
\begin{eqnarray}
F(R) \sim \bar{F} R^{7/2}\,
\label{eq:3.12}
\end{eqnarray}
for $t_s^2R \gg 1$,
where
\begin{eqnarray}
\bar{F} = 
\frac{2}{7}
\left[
\frac{1}{3\sqrt{140} \left( 2\gamma +1 \right)}
\left( \frac{t_s}{t_0} \right)^\gamma \right]^5
\left(
\sum_{j=\pm}
\tilde{p}_j t_s^{\beta_j} \right)
t_s^5\,.
\label{eq:3.13}
\end{eqnarray} 
In this case, we obtain
\begin{eqnarray}
J = \frac{441000 \left(72-71b\right) \bar{F}}{\left(t_s-t\right)^6}
R^{3/2}\,.
\label{eq:3.14}
\end{eqnarray} 
For $\gamma > 1/20$, because $R>0$ and $0<b<1$, 
the necessary condition to have $J \geq 0$ in Eq.~(\ref{eq:3.14}) is 
\begin{eqnarray}
\sum_{j=\pm} \left( \tilde{p}_j t_s^{\beta_j} \right) \geq 0\,,
\label{eq:3.15}
\end{eqnarray} 
which can be met by choosing the 
appropriate values of $\tilde{p}_\pm$.
Since $\tilde{p}_\pm$ are arbitrary integration 
constants, for simplicity,
we choose $\tilde{p}_+ = 0$.
When $\gamma \sim {\mathcal O}(1)$ and 
$\tilde{p}_- > 0$, $F(R)$ is always  positive in both  non-phantom and 
phantom phases. This is reasonable because for general relativity, 
$F(R) = R >0$. 
Consequently, in this $F(R)$ gravity model with the crossing of the phantom 
divide, the second law of thermodynamics can be satisfied. 

For a power-low type $F(R)$ gravity described as 
$F(R) = c_1 M^2 \left( R/M^2 \right)^{-n}$, 
where $c_1$ and $n$ are dimensionless constants and $M$ denotes a mass scale, 
the scale factor $a(t)$ is given by 
$a(t) = \bar{a} \left( t_s - t \right)^{(n+1)(2n+1)/(n+2)}
$~\cite{Bamba:2008hq, Briscese:2006xu}. 
The form of $F(R)$ in Eq.~(\ref{eq:3.12}) corresponds to the case with 
$n=-7/2$. 
Accordingly, $a(t)=\bar{a} \left( t_s - t \right)^{-10}$, 
which implies $\ddot{a} = 110\bar{a} \left( t_s - t \right)^{-12} > 0$. 
Thus, a late-time cosmic acceleration 
can be realized. This is the outcome of the $F(R)$ gravity model 
in Eq.~(\ref{eq:3.12}). 
It should be noted that whether the gravity model of 
$F(R) \sim \bar{F} R^{7/2}$ in Eq.~(\ref{eq:3.12}) can pass Solar System tests 
still needs to be examined~\cite{Chiba:2006jp}. 

 From Eq.~(\ref{eq:3.10}), we obtain $R=1260/\left(t_s-t\right)^2$, which 
leads to 
$J = \left[ \left(72-71b\right) \bar{F}/4536 \right] R^{9/2}$ based on 
Eq.~(\ref{eq:3.14}). In the thermal equilibrium 
limit, i.e., $b \sim 1$, we find 
$J \sim \left(\bar{F}/4536 \right) R^{9/2}$. 
On the other hand, for general relativity with $F(R)=R$, it follows from 
Eq.~(\ref{eq:2.40}) with $b=1$ that $J=\dot{H}^2 = 
\left(1/15876 \right) R^2$ by 
assuming the same behavior of $H$ in Eq.~(\ref{eq:3.10}). 
The main difference between the expressions of $J$ in the thermal equilibrium 
limit and that for general relativity is only the power of $R$. 
This comes from the difference of the action between the present $F(R)$ 
gravity $F(R) \sim \bar{F} R^{7/2}$ in Eq.~(\ref{eq:3.12}) and general 
relativity with $F(R)=R$. 

Finally, we remark that even in the effective phantom era of this $F(R)$ 
gravity model, the second law of thermodynamics can be satisfied due to the 
non-equilibrium thermodynamic treatment. Hence, this model is more similar to 
a phantom model with ordinary thermodynamics suggested in 
Ref.~\cite{Nojiri:2005sr}.

\section{Conclusion}

We have investigated the first and second laws of 
thermodynamics of the apparent horizon in $F(R)$ gravity~\cite{Wu:2008ir}. 
We have shown that in the $F(R)$ gravity model with realizing the crossing 
of the phantom divide proposed in Ref.~\cite{Bamba:2008hq}, the second law of 
thermodynamics can be satisfied in not only the non-phantom phase but also the 
effective phantom one. In addition to cosmological constraints and 
solar system tests on the models of $F(R)$ gravity, such an examination 
whether the second law of thermodynamics can be met in those models is 
important. The demonstration in this work can be regarded as a meaningful 
step to construct a more realistic model of $F(R)$ gravity, which could 
correctly describe the expansion history of the universe.

\section*{Acknowledgments}
We thank Professor Sergei D. Odintsov and Professor Shin'ichi Nojiri for their 
collaboration in our previous work~\cite{Bamba:2008hq} and important comments. 
We are grateful to Professor Shinji Tsujikawa for very helpful 
discussions. 
We also appreciate communications with Dr. Tsutomu Kobayashi. 
This work is supported in part by
the National Science Council of R.O.C. under:
Grant \#s: NSC-95-2112-M-007-059-MY3 and
National Tsing Hua University under Grant \#:
97N2309F1 (NTHU).


\end{document}